\documentclass[11pt]{article}
\usepackage{epsfig}
\usepackage{epsf}
\newlength{\xfigsize}
\newlength{\halffigsize}
\newcommand{\fig}[5]                    
                {
                \begin{figure}
                        \epsfclipon
                        \setlength{\epsfxsize}{#1}      
                        \setlength{\epsfysize}{#2}      
                        \hspace*{\fill} \epsffile#3 \hspace*{\fill}
                        \caption{#4}
                        \label{#5}
                \end{figure}
                }

\begin{document}

\begin{titlepage}

\vskip 3cm

\begin{centering}

{\huge
The ZEUS Forward Plug Calorimeter with}

{\huge 
Lead-Scintillator Plates and WLS Fiber}

\vspace*{0.2 cm}
{\huge
Readout}

\vskip 1cm

{\Large The ZEUS FPC Group}

\end{centering}

\vskip 1cm

{\rm \normalsize

A. Bamberger$^2$, S. B\"ottcher$^{8,13}$, I. Bohnet$^3$, 
J.P. Fern\'andez$^{6,12}$, F. Goebel$^1$, P. G\"ottlicher$^1$, 
A. Gabareen$^8$, G. Garc\'{\i}a$^{6,*}$, N. Gendner$^3$, R. Graciani$^1$, 
M. Hauser$^{2,11}$, D. Horstmann$^1$, M.~Inuzuka$^9$, 
M. Kasemann$^{1,12}$, B. L\"ohr$^1$, R. Lewis$^7$, 
H. Lim$^5$, L. Lindemann$^{1,16}$, P. Markun$^2$, 
M. Mart\'{\i}nez$^1$, T. Neumann$^{3,14}$, I.H. Park$^5$, 
J. del Peso$^{6,15}$, H. Raach$^2$, A. Savin$^1$, D. Son$^5$,
K.~Tokushuku$^4$, S. W\"olfle$^2$, J. Whitmore$^7$,
K. Wick$^3$, G. Wolf$^1$, S.~Yamada$^4$, 
T.~Yamashita$^{10}$, Y.~Yamazaki$^4$

\vskip 1cm

{\it \small

$^1$  Deutsches Elektronen-Synchrotron DESY, Hamburg, Germany \\
$^2$  Fakult\"at f\"ur Physik der Universit\"at Freiburg, Freiburg,
Germany $^a$ \\
$^3$  Hamburg University, I. Institute of Exp. Physics, Hamburg, Germany $^a$\\
$^4$  Institute of Particle and Nuclear Studies, KEK, Tsukuba, Japan\\
$^5$  Kyungpook National University, Taegu, Korea $^b$\\
$^6$  Univer. Aut\'onoma Madrid, Dpto de F\'\i sica Te\'orica, Madrid,
 Spain $^c$ \\
$^7$  Pennsylvania State University, Dept. of Physics, University Park,
PA, USA $^d$\\
$^8$  School of Physics, Tel-Aviv University, Tel Aviv, Israel $^e$\\
$^9$  Tokyo Metropolitan University, Dept. of Physics, Tokyo, Japan $^f$\\
$^{10}$ Department of Physics, University of Tokyo, Tokyo, Japan $^f$\\
$^{11}$ now at University of Bern, Switzerland\\
$^{12}$ now at Fermilab, Batavia, USA\\
$^{13}$ now at Nevis Labs., N.Y., USA \\
$^{14}$ now at Deutsche Bank, Frankfurt/M, Germany \\
$^{15}$ partially supported by Comunidad de Madrid \\
$^{16}$ now at SAP, Walldorf, Germany \\

$^a$ supported by the German Federal Ministry for Education and Science,
Research and Technology (BMBF) \\

\newpage

$^b$ supported by the Korean Ministry of Education and Korea Science and 
Engineering Foundation \\
$^c$ supported by the Spanish Ministry of Education and Science
through funds provided by CICYT \\
$^d$ supported by the US National Science Foundation \\
$^e$ supported by the Israeli Science Foundation \\
$^f$ supported by the Japanese Ministry of Education, Science and Culture \\

$^*$ Corresponding author. Tel: 49-40-89983752; fax: 49-40-89983092 ; \\
e-mail: garcia@mail.desy.de \\
}

}

\begin{centering}


\vskip 1cm


\vspace{2cm}

\begin{abstract}
A Forward Plug Calorimeter (FPC) for the ZEUS detector at HERA
has been built as a shashlik
lead-scintillator 
calorimeter with wave length shifter fiber readout.
Before installation it was
tested and calibrated using the X5 test beam facility of the SPS
accelerator at CERN.
Electron, muon and pion beams in the momentum range of 10 to 100 GeV/c
were used.
Results of these measurements are presented as well as a calibration 
monitoring system based on a $^{60}$Co source.
\end{abstract}

\end{centering}

\end{titlepage}

\newpage
\clearpage

\section{Introduction}

The ZEUS \cite{kn:bluebook} collaboration has installed a Forward Plug
Calorimeter (FPC) around the beam line
(see Fig.~\ref{fig:fpcinzeus}) to extend the calorimetric coverage
in pseudorapidity~\footnote{Pseudorapidity is defined as 
$\eta = -\ln \left(\tan (\theta/2)\right)$, where $\theta$ stands for
the angle between the particle trajectory and the forward
proton beam direction.}
from $\eta \leq 4.0$ to $\eta \leq 5.0$.
This vastly increases the physics potential for diffraction in deep 
inelastic scattering.
The mass range over which the dissociated photon system can be studied
is enhanced by a factor of about two \cite{kn:FPCprop}.

This document describes the construction and 
the results of the beam test
of the FPC performed at CERN in
September 1997 with electrons, muons and pions covering an energy range
from 10 to 100 GeV. 
Prior to this,
a prototype corresponding to one 
half of the electromagnetic section was tested at
DESY with electron beams from 1 to 6 GeV \cite{kn:DESYtestbeam}.
Data from the beam test at DESY
will also be presented where appropriate.

\section{Description of the detector}
\label{sec-description}

\begin{figure}
\setlength{\epsfxsize}{8cm}
\centerline{
\hspace*{2.0 cm}\mbox{\epsfig{figure=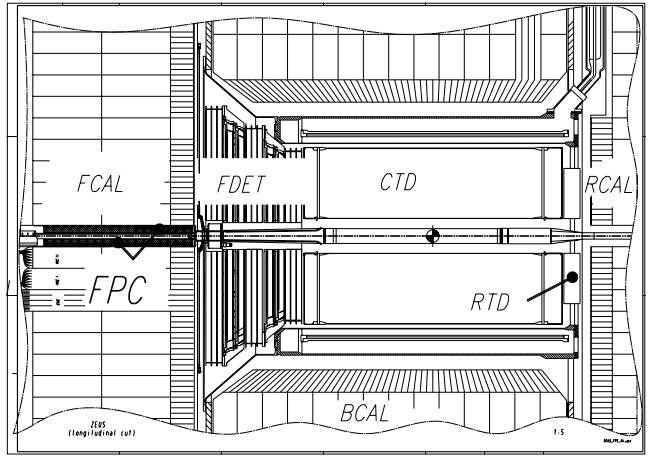,angle=90,width=6.0 cm}}}
\vspace*{0.0 in}
\caption{\em Side view of FPC integrated in the ZEUS Calorimeter.
}
\label{fig:fpcinzeus}
\end{figure}

\begin{figure}
\setlength{\epsfxsize}{8cm}
\centerline{\mbox{\epsffile{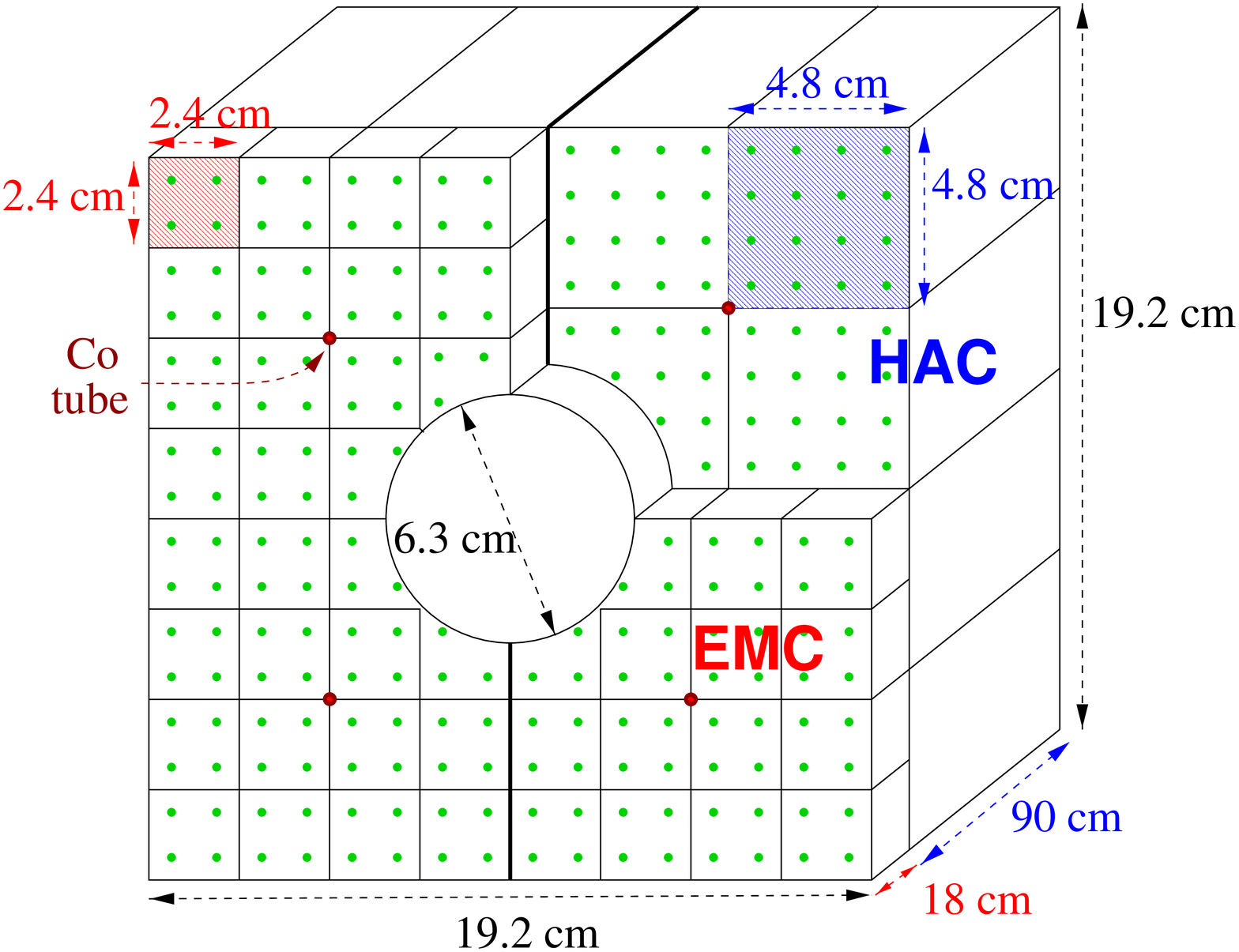}}}
\vspace*{0.0 in}
\caption{\em Front view of the FPC.
The readout cells and the position of WLS fibers are shown. Notice
that there is one hadronic readout cell behind $2 \times 2$
electromagnetic cells, except for the cells near the beam hole.}
\label{fig:f_zeus03}
\end{figure}

The FPC is a lead-scintillator sandwich calorimeter read out by
wave length shifter
(WLS) fibers and photomultipliers (PMT).
This concept has been investigated 
in~\cite{kn:calfibras1,kn:calfibras2,kn:calfibras3}.
It has been installed in the $20 \times 20\ $cm$^2$ beam
hole of the forward uranium-scintillator 
calorimeter (FCAL)
of the ZEUS
detector at HERA ~\cite{kn:bluebook}
(see Fig.~\ref{fig:fpcinzeus}).
A front view of the FPC is shown in
 Fig.~\ref{fig:f_zeus03}. 
 The FPC has a 63 mm
diameter
central hole to accommodate the HERA beampipe.

The active part of the FPC has outer dimensions of
$192 \times 192 \times 1080$ mm$^3$. The FPC 
is mechanically subdivided
into two identical half modules. They are attached
to the bottom and top half parts of the innermost FCAL
module~\cite{kn:FPCprop}. The FCAL
halves and therefore also the FPC halves are moved apart
for beam injection in order to reduce the radiation
dose on the calorimeters.

In the FPC,
lead plates of $15\ $mm thickness alternate with scintillator layers of
$2.6\ $mm.
The WLS fibers have $1.2\ $mm diameter and pass through $1.4\ $mm
diameter holes in the lead and scintillator layers.
The holes are located
on a $12\ $mm step square grid.
The FPC has 232 holes of this type.
In addition, there are 4 brass tubes 
($1.4\ $mm inner diameter)
which are used to guide
a $^{60}$Co source, placed on the tip of a long steel wire,
for the monitoring of the calibration of the FPC cells.
Taking into account the WLS fibers and the fiber holes in
the lead plates, the effective lead to plastic ratio by volume
is 5.2:1. Based on the results from a lead-scintillator calorimeter
of similar composition~\cite{kn:perf}
the FPC is expected to provide equal response to electrons and hadrons
(compensating calorimeter, $e/h = 1$).

With the layer structure chosen, the FPC has approximately
the same radiation length $X_0$ and nuclear absorption length $\lambda$
as the FCAL, viz. 
$X_0($FPC$) = 0.68\ $cm and $X_0($FCAL$) = 0.74\ $cm;
$\lambda($FPC$) = 
20 \ $cm and $\lambda($FCAL$) = 21.0\ $cm.
This minimizes the fluctuations in the 
energy measurement.

\begin{figure}
\begin{center}
\setlength{\epsfxsize}{12cm}
\centerline{\mbox{\epsffile{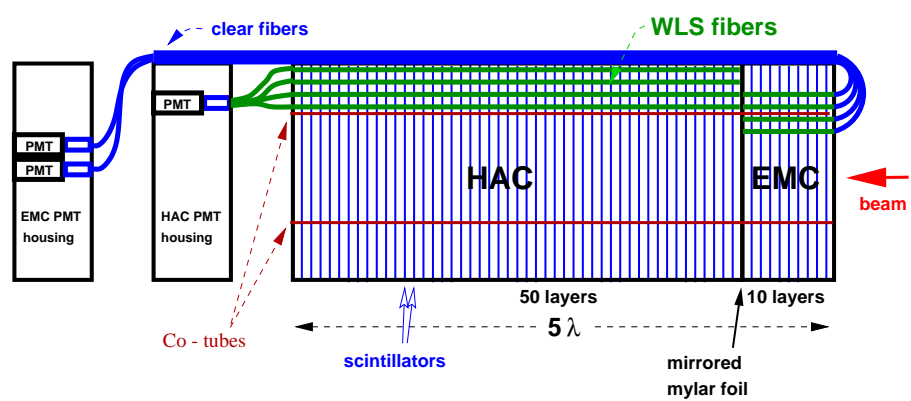}}}
\vspace*{0.0 in}
\caption{\em Schematic side view of the FPC.
The horizontal lines indicate the positions of the WLS fibers
and of the tubes for the $^{60}$Co source.}
\label{fig:f_zeus04}
\end{center}
\end{figure}

The FPC is subdivided longitudinally into an electromagnetic (EMC) and
a hadronic (HAC) section which are read out separately
(see Figs.~\ref{fig:f_zeus03} and~\ref{fig:f_zeus04}).
The electromagnetic section consists of 10 layers of lead and scintillator
corresponding to $26.5 \, X_0$ and
$0.9 \, \lambda$.
The hadronic section of the FPC consists of 50 layers and represents
$4.5 \ \lambda$ leading to a total for the FPC of $5.4\ \lambda$
(see Table~\ref{tab:FPCparameters}).

The scintillator layers consist of tiles and form cells which are
read out individually (Fig.~\ref{fig:f_zeus03}).
The cell cross sections are $24 \times 24\ $mm$^2$ in the EMC,
commensurate with the Moliere radius, and
$48 \times 48\ $mm$^2$ in the HAC section.
The 8 (4) innermost cells in EMC (HAC) surrounding the beam hole
follow the circular shape given by the beam hole.

The polystyrene based scintillator SCSN81T2
from Kuraray was used, since it was found to 
 be the best choice in terms of light
yield and radiation stability ~\cite{kn:scintillator}.
The scintillator tiles were wrapped with $0.2\ $mm thick tyvek paper 
in order to improve the light collection efficiency
and avoid light coupling between neighboring cells.

The WLS fibers of the EMC are connected to clear fibers ($1.4\ $m length)
which transport the light to the PMTs placed behind the FPC
(see Fig.~\ref{fig:f_zeus04}).
The clear fibers are bent by $180^\circ$ at the front of the EMC,
and are guided to the PMTs in the rear by two ducts situated 
on either side
of the FPC.
The WLS fibers of the HAC section transport the light directly to 
the PMTs.
All 4
(16) fibers corresponding to an EMC (HAC) cell are
connected to the same PMT through a light-mixer bar.
An additional
fiber is connected to each PMT through 
the same light-mixer bar
in order to inject LED and laser light pulses  for monitoring
the stability of the PMTs and readout
electronics.
On the other end of the WLS fibers a 
reflective aluminized mylar foil
 is placed to 
avoid light losses (see Fig.~\ref{fig:f_zeus04}).
The total number of readout channels is 
(EMC + HAC): 60 + 16 = 76 (see Table~\ref{tab:FPCparameters}).

For the WLS fibers the material Y11200(dc) 
from Kuraray has been chosen. It has
a polystyrene core and a double cladding (PMMA and fluorinated
PMMA) which produces a substantial increase in light output
compared to a single clad fiber. Its
absorption spectrum matches best the SCSN81T2 emission spectrum.

The PMT chosen for the FPC is the Hamamatsu R5600U, a
tube with metal channel dynodes. It is 
relatively insensitive to
 magnetic fields and its
small dimensions ($16 \times 16 \times 16\ $mm$^3$) are 
well suited for the limited
space available around the hole of the FCAL.
 Tests done  prior to the
beam tests showed that it fulfills well 
the constraints on dynamic ranges placed by the
required FPC performance in the HERA 
environment~\cite{kn:FPCprop}.

The PMTs are placed in four separate iron blocks
whose shape and
transverse size is similar to that of the lead plates.
The four structures correspond to the EMC and HAC 
sections of both halves.
 These blocks are placed
behind the hadronic section, first the one containing
the HAC PMTs and 
then that for the EMC
(see Fig.~\ref{fig:f_zeus04}).
The EMC PMTs are arranged in their support frames in such
a way that they do not line up with their respective 
calorimeter cell.

After the installation in ZEUS, the PMTs are connected to a high
voltage system based on a Cockroft-Walton generator ~\cite{kn:hvsystem}.

\begin{table}[t]
\begin{center}
\begin{tabular}{|c|c|} \hline
 
unit layer                   & $15 \, $mm Pb, $2.6 \, $mm scint, $0.4 \,
$mm tyvek paper \\
effective thickness of layer & $14.84 \, $mm Pb, $2.86 \, $mm plastic \\
\hline
transversal dimensions       & $192 \times 192 \, $mm$^2$              \\
EMC section, 10 layers       & $180 \, $mm
\\
HAC section, 50 layers       & $900 \, $mm
\\
diameter beam hole           & $63 \, $mm                               \\
\hline
number of cells              &                                     \\
EMC $+$ HAC                  & 60 $+$ 16                           \\
\hline
average density              & $9.6 \, $g/cm$^3$                        \\
radiation length ($X_0$)     & $0.68 \, $cm                             \\
absorption length ($\lambda$)& $20 \, $cm                             \\
Moliere radius               & $2.0 \, $cm                              \\
\hline
total length                 & $1080 \, $mm                             \\
\hline
total weight                 & $400 \, $kg                              \\
\hline
total radiation length:      &                                     \\
EMC $+$ HAC                  & 26.5 $+$ 133.0 $=$ 159.5 $X_0$         \\
\hline
total absorption length:     &                                     \\
EMC $+$ HAC                  & 0.9 $+$ 4.5 $=$ 5.4 $\lambda$    \\
\hline
 
\end{tabular}
\caption{\em Summary of FPC parameters.}
\label{tab:FPCparameters}
\end{center}
\end{table}

\section{CERN setup}
\label{sec-setup}

\begin{figure}
\begin{center}
\setlength{\epsfxsize}{8cm}
\centerline{\mbox{\epsffile{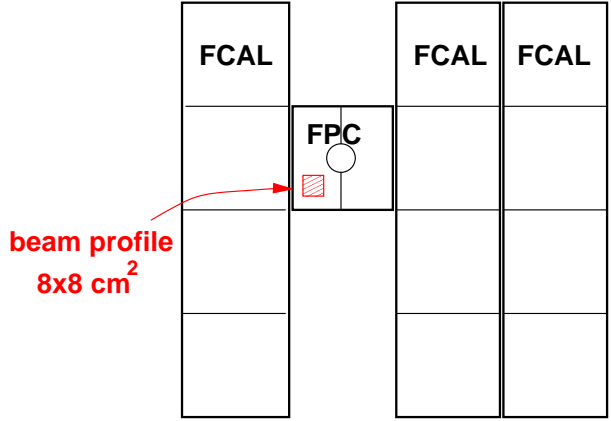}}}
\vspace*{0.0 in}
\caption{\em Front view of the FPC at the test beam site, surrounded by
three FCAL modules.}
\label{fig:cern_engl}
\end{center}
\end{figure}

The FPC has been tested in the X5 beam of the CERN SPS.
 It has been installed
between modules of the FCAL prototype,
 see Fig.~\ref{fig:cern_engl}.
These are uranium-scintillator calorimeter units similar
to the ones implemented at ZEUS (see ~\cite{kn:FCAL}).
The uranium plates have a thickness of $3.3\ $mm and 
the scintillator layers are $2.6\ $mm thick.
The FCAL units are divided longitudinally into 3 sections,
one electromagnetic, $26 \, X_0$, and 2 hadronic sections,
3 $\lambda$ each.
The electromagnetic (hadronic) sections are 
divided transversally into $5 \times 20\ $cm$^2$  cells
($20 \times 20\ $cm$^2$ cells).

The combined FPC $+$ FCAL prototype setup has been 
placed on a structure, 
which could be moved
in the $x$ and $y$ directions, in order to
vary the impact point of the particle beam at the calorimeter.
The beam line defines the $z$-direction.

\begin{figure}
\begin{center}
\setlength{\epsfxsize}{12cm}
\centerline{\mbox{\epsffile{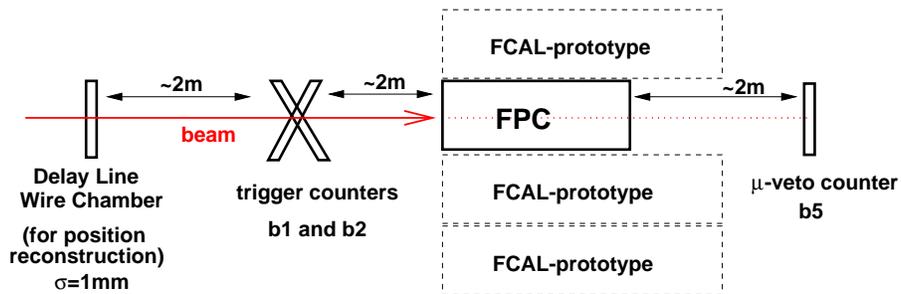}}}
\vspace*{0.0 in}
\caption{\em Top view sketch of the beam test setup.}
\label{fig:cerntest_engl}
\end{center}
\end{figure}

 A Delay Line Wire Chamber (DLWC)
in front of the FPC
has given
precise  information on the charged particle position
(see Fig.~\ref{fig:cerntest_engl}).
 The
relative position of the FPC detector with respect to the
DLWC coordinates
has been 
 established to within about
one millimeter.
Cuts
on DLWC coordinates were used in the offline analysis
in order to select particles
hitting a given region of the detector. 

A defocused beam was used to obtain events covering a wide region
of the FPC. The trigger was defined by requiring signals in both
scintillator counters ($b_1$ and $b_2$) placed in front of the FPC.
Behind the FPC an additional counter ($b_5$) 
was installed mainly for offline identification of muons.

\section{Simulation of the FPC measurements performed at CERN}

For the analysis of the CERN-test data the FPC has been simulated
by Monte Carlo (MC)
using the GEANT 3.21 package \cite{kn:geant312}. 
The FPC has been implemented as a sampling
calorimeter consisting of scintillator and absorber layers. The density
of the absorber layers has been reduced 
in order to
account for the tyvek layer
and for the holes left for the WLS fibers.
 The fibers were not included in the simulation.
The light attenuation in the WLS fibers has
been taken into account according to the measurement described in
section~\ref{att-length}. Photostatistics, noise and
cross talk have also been included according to measurements results
(see section~\ref{crosstalk}).
In the simulation the
FPC has been positioned between FCAL modules as in the CERN-test
setup described in the previous section (see also Fig.~\ref{fig:cern_engl}). 

The accuracy of the simulated response to electrons and muons has been
estimated to be about 3\%. 
For pions the accuracy of the
simulated 
energy signals, which have been determined with 
the hadronic package GHEISHA \cite{kn:gheisha}, is about $\pm$10\%.

\section{Calibration}

For the calibration of the FPC the following scheme has been chosen:

\begin{itemize}
\item calibration of EMC cells using electron test beam data,
\item calibration of HAC section using muon test beam data,
\item 
 monitoring of the calibration constants using the
  response of each FPC channel to irradiation with a $^{60}$Co source.
\end{itemize}

\subsection{Calibration of the EMC section}
\label{calib_emc}

A data set of 60 GeV test beam electrons distributed over the complete
FPC surface has been used for the calibration of the EMC cells. 
The
calibration constants have been adjusted such that the mean of the
summed signal of all EMC cells is independent of the point of
incidence. The absolute calibration has been adjusted to the beam energy.
For the edge cells the beam energy is reduced by a correction factor,
estimated by MC,
taking into account the energy leaking out of the FPC.

\subsection{Calibration of the HAC section}

\begin{figure}
\begin{center}
\setlength{\epsfxsize}{12cm}
\epsffile{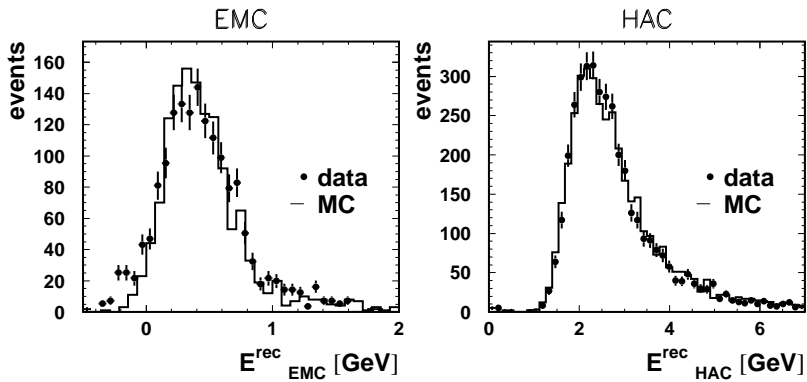}
\vspace*{-0.0 cm}
\caption{\em The reconstructed energy 
measured in the EMC and HAC sections of the FPC
for 100 GeV muons is shown
  for data and MC.}
\label{fig:muon_hac_plots}
\end{center}
\end{figure}

Since the energy  of pions is not fully contained in the FPC
and the amount of leakage is not reliably described by the MC,
 muons have been used for the calibration of the HAC cells.  
A Landau function convoluted with a gaussian function, to take into
account photostatistics and electronic noise, has been fitted to the
muon signals. 
The calibration constants have been
computed in order to adjust the peak value obtained from the fit, to
the value predicted by MC.
Since the distribution of the muon signals does not have
a gaussian shape,
it is crucial that it is well described by MC.
A good overall agreement is observed for the HAC section
after calibrating with muons, see
Fig.~\ref{fig:muon_hac_plots}.

In order to measure hadronic energy,
the following
two corrections have been applied:

\begin{itemize}
\item 
The ratio between the visible and the total energy deposited in
the FPC (sampling fraction) is different for an incident pion and
an incident muon.
Assuming
that the FPC is a compensating
calorimeter ($e/h = 1$),
the sampling fraction obtained with electrons (see 
section~\ref{calib_emc})
can also
be used for pions.
\item 
The effect of the light attenuation in the
WLS fibers (see section~\ref{att-length})
 is different for muons and pions due to their
  different 
longitudinal energy deposition.
 Whereas muons deposit energy
uniformly along the $z$-direction, pions deposit  their energy 
predominantly at the beginning of the
 HAC region where the effect of the light attenuation is strongest.
In order to correct for the effect, the energy scale of the HAC section
has been raised by a factor of 1.33.
\end{itemize}

The accuracy of the absolute calibration thus obtained is estimated to
be 5\%. The error is dominated by the
uncertainty of the MC prediction.

\subsection{Cross talk}
\label{crosstalk}

A cross talk effect has been observed between
physically adjacent cells of the FPC
in the analysis of the CERN-test data.
No  cross talk  between
 PMTs or front-end 
electronic channels
has been found. The observed cross talk has
therefore been attributed to light being produced in one cell, crossing the
tyvek barrier between scintillator tiles and entering the
WLS fibers of a neighboring cell. It is known that
tyvek is not totally light tight. 

The cross talk has been  measured using the  muon sample. 
For all pairs of neighboring cells the ratio between the signal of
the cell that has been hit and the signal in the adjacent cell has
been determined. The mean value of this distribution has been taken as
the amount of cross talk.

For light produced in the scintillator of a given cell
 (denoted as the central cell) the
fraction of light that
has been collected in this and the adjacent cells is shown 
in Table~\ref{tab:crosstalk}.

\begin{table}
\begin{center}
\begin{tabular}{|l|r|r|} \hline
                                    & EMC   & HAC   \\ \hline
central cell                  & 73\%  &  81\% \\ \hline
each of 4 direct neighbors    & 5.8\% & 4.3\% \\ \hline
each of 4 diagonal neighbors &   1\% & 0.4\% \\ \hline
\end{tabular}
\caption{\em Fraction of light collected in the central cell
and in the adjacent ones.}
\label{tab:crosstalk}
\end{center}
\end{table}

Note that the method applied to calibrate the EMC (HAC), which has
been described in previous sections, is not sensitive to cross talk,
since it uses the sum of the signals over all cells.

In order to describe the lateral shower width,
the measured cross talk has been implemented in the MC.

\subsection{$^{60}$Co monitoring system}
\label{sec-cobalt}

A monitor system using a $^{60}$Co source, similar to the one used for
the ZEUS uranium calorimeter~\cite{kn:Co1},
 has been developed for the FPC. It
allows the detection of changes in the performance of the scintillator tiles
and the WLS fibers as well as drifts in the gain of the PMTs. By
measuring the ratio of response to $^{60}$Co and beam particles
the absolute and cell-to-cell 
calibration constants can be transported from the test beam
to ZEUS and the stability of the calibration can be monitored.

\subsubsection{$^{60}$Co setup}

\begin{figure}
\begin{center}
\setlength{\epsfxsize}{12cm}
\epsffile{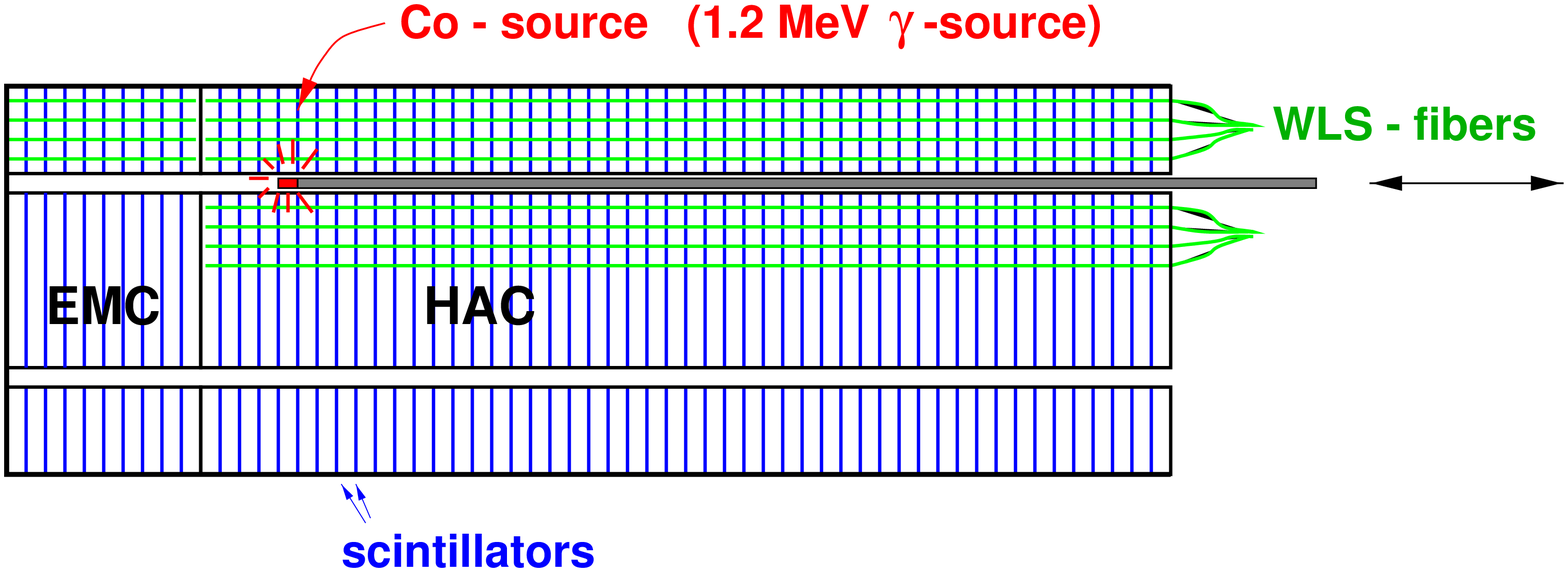}
\end{center}
\vspace*{-0.5 cm}
\caption{\em Schematic drawing of the FPC with inserted $^{60}$Co source}
\label{fig:cobalt-drawing}
\end{figure}

A 1 mCi pointlike $^{60}$Co source is attached to the tip of a steel
wire. For safety reasons the wire and the source are enclosed inside a
stainless-steel tube. This source wire can be inserted into brass
tubes of $1.4\ $mm inner diameter inside the FPC. The brass tubes run
 parallel to the WLS fibers at the center of each FPC quarter
(see Fig.~\ref{fig:f_zeus03}).  The $^{60}$Co source
irradiates the scintillators of the FPC with 1.173 MeV and 1.332 MeV
photons (see Fig.~\ref{fig:cobalt-drawing}).

A PC-controlled motor moves the $^{60}$Co source wire in steps of $0.6
\ $mm through the FPC. At each step the PMT currents are read out across
2~M$\Omega$ resistors by an integrating, voltage sensitive ADC. The
measurement is repeated 500 times and the mean and 
root mean square (RMS) values are
recorded. During a complete scan the source fully traverses the FPC
(from EMC to HAC) and the 15 EMC and 4 HAC channels of the
corresponding FPC quarter are read out.

\begin{figure}
\begin{center}
\setlength{\epsfxsize}{10cm}
\epsffile{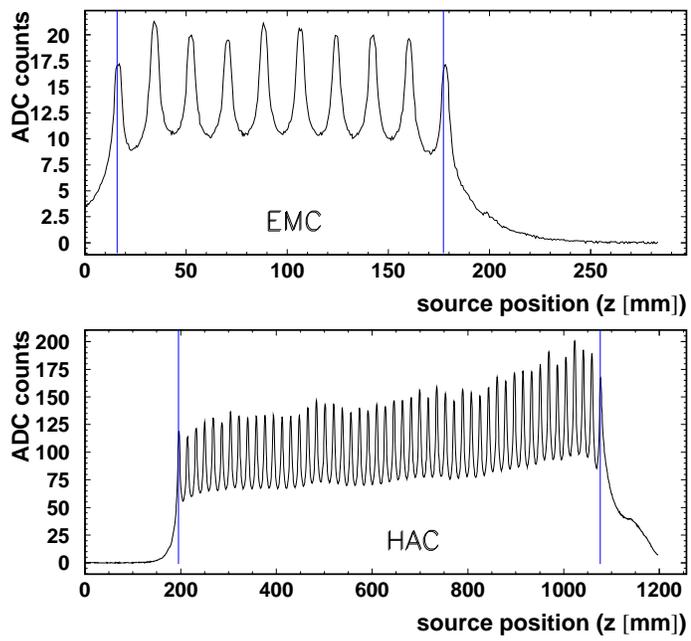}
\end{center}
\vspace*{-1.0 cm}
\caption{\em 
The measured
 $^{60}$Co signal
in the EMC and HAC sections
 as a function of
  the source position. The vertical lines indicate the
  region of integration 
used to obtain the value of COMEAN.}
\label{fig:cobalt-scan}
\end{figure}

The signal as a function of position is shown in Fig.~\ref{fig:cobalt-scan}. 
The 10 (50) peaks in the EMC (HAC) part
correspond to positions where the source is next to a scintillator.
When the source is inside a lead layer the photons are partially
shielded and the signal drops. 
The point where
 the
source exits the EMC section and enters the HAC section
is clearly seen at
  $z \sim
190\ $mm.

The maxima of the signals from the individual scintillator tiles
fluctuate.
This is assumed
to be due 
to differences in the scintillator-WLS fiber light coupling and
to
differences in the tyvek wrapping and scintillator 
machining. The overall
increase of the signal in the HAC section 
with increasing $z$
is due to light
attenuation  in the WLS fibers: the fibers are read
out in the direction of positive $z$ (see section~\ref{att-length}).

\subsubsection{Monitoring of the stability of the signal}
\label{co-stabil}

\begin{figure}
\begin{center}
\setlength{\epsfxsize}{10cm}
\epsffile{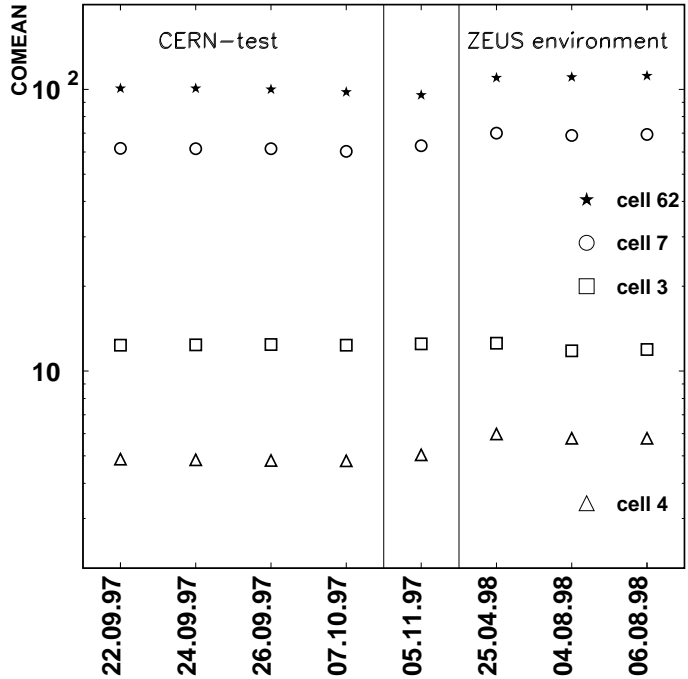}
\end{center}
\caption{\em The COMEAN values from different $^{60}Co$ scans,
listed by dates, are shown
  for HAC cell 62 and for three EMC cells at different distances to the
  cobalt source. According to the classification in Fig.~\ref{fig:co-geom}
  cell 7 belongs to group A, cell 3 to B and cell 4 to C.}
\label{fig:comean}
\end{figure}

In order
to study the stability of the FPC response, the signal induced by the
$^{60}$Co source
is integrated between the bounds shown in Fig.~\ref{fig:cobalt-scan}.
The result of the integration is called COMEAN.
Figure~\ref{fig:comean} shows COMEAN for 8 different $^{60}$Co scans 
spanning the time
from the
CERN-test period in September 1997 
until the installation of the FPC in ZEUS and 
the start of luminosity runs in August 1998.
Up to 10\% deviations can be observed after the transportation
of the FPC
to DESY (05.11.97) and after installation of the FPC into the ZEUS
environment (25.04.98). 

\begin{figure}
\begin{center}
\setlength{\epsfxsize}{8cm}
\epsffile{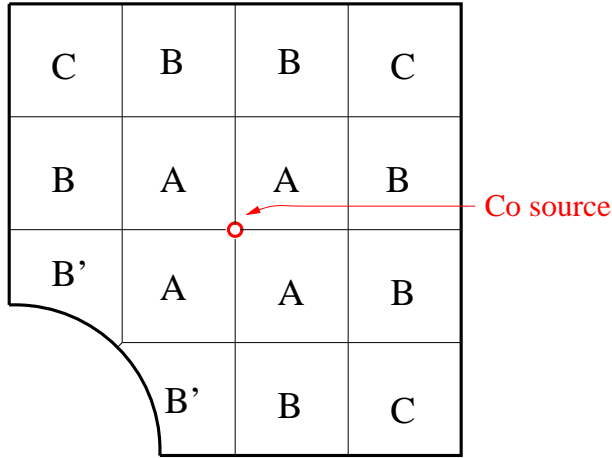}
\end{center}
\caption{\em Front view of a quarter of the FPC. The EMC cells are
  classified in three groups depending on their distance to the $^{60}Co$
  source.}
\label{fig:co-geom}
\end{figure}

The different signal heights observed for the different cells 
 in
Fig.~\ref{fig:comean} is due to their distances
from  the cobalt source.
The EMC cells can be subdivided into 3 groups (A,B,C)
depending on their distance from the $^{60}$Co source (see
 Fig.~\ref{fig:co-geom}).
The distance between the $^{60}$Co source and HAC cells is
 the same for all HAC cells.

\subsubsection{Measurement of the light attenuation in WLS fibers}
\label{att-length}

\begin{figure}
\begin{center}
\setlength{\epsfxsize}{13cm}
\epsffile{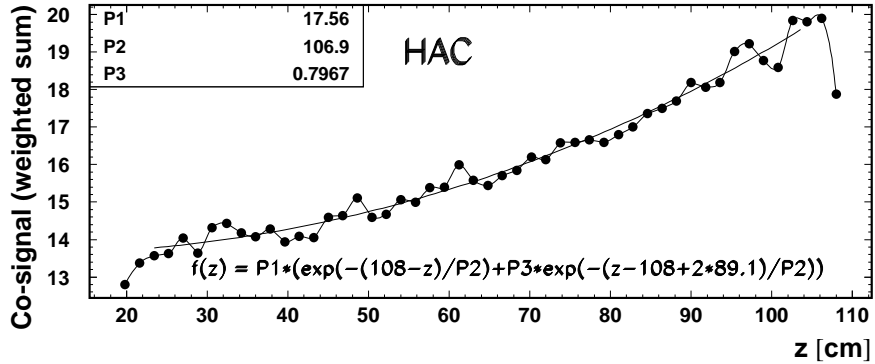}
\end{center}
\vspace*{-1.0 cm}
\caption{\em The peak height averaged over all cells is plotted as a
  function of peak position $z$. The attenuation length in the
  WLS fibers is determined with a two exponential fit. The second
  exponential accounts for the light reflected at the end of the
  fiber.}
\label{fig:att-length}
\end{figure}

The change of the peak height with the source position $z$ 
in the HAC section
(see Fig.~\ref{fig:cobalt-scan})
 allows the measurement of the light attenuation length in the
WLS fibers directly. Since the peak height is also influenced by the
scintillator quality and the light coupling into the WLS fiber, average
peak values of all channels have been used (see Fig.~\ref{fig:att-length}).

The data are fitted by using the function:

    \begin{equation}
      f(z) = P_1 \cdot (e^{- (z_0 - z) / P_2} + P_3 \cdot e^{-(z - z_0 + 2
\cdot L) / P_2}) \label{eq:attl}
    \end{equation}

      where  $L  =  89.1\ $cm  is the length of the HAC 
            (distance from the
             reflective mylar foil to the last scintillator layer on
             the back) and $z_0 = 108.0\ $cm is the z-position where the
             fiber exits the FPC.
In Eq.~\ref{eq:attl}
the second exponential accounts for the light reflected at the mylar
foil which separates the EMC and HAC sections;
$P_1$, $P_2$ and $P_3$ are free parameters in the fit.
The attenuation length measured using this method is
$P_2 = 107\ $cm.
The weight of the second exponential function is
$P_3 = 0.80$,
which indicates that 80\% of the light is reflected by the
mylar foil.

\section{Effect of WLS fibers on the signal}

\subsection{Signal uniformity}
\label{sec-uniformity}

The uniformity of the response across the surface of a FPC cell is
affected by the presence of the WLS fibers. Light produced in
scintillator regions close to a WLS fiber is collected with a better
efficiency than in other regions as discussed in \cite{kn:calfibras2}.

\begin{figure}
\begin{center}
\setlength{\epsfxsize}{13cm}
\hspace*{-2cm}
\epsffile{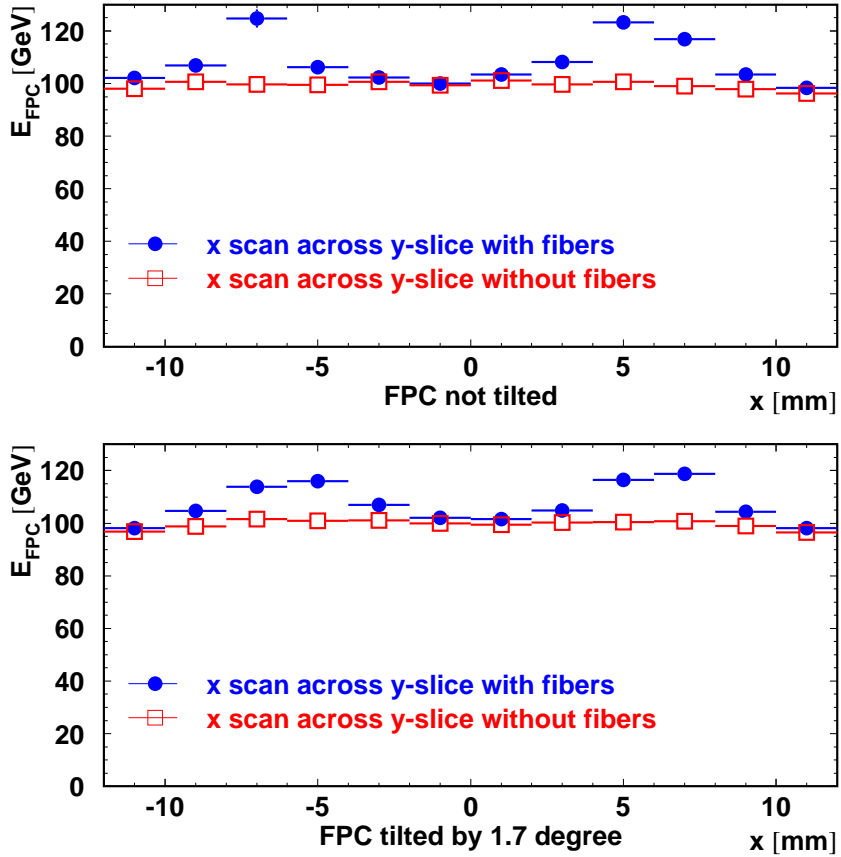}
\end{center}
\vspace*{-0.9 cm}
\caption{\em 
The energy measured by the FPC,
$E_{FPC}$, for 100 GeV electrons is plotted as a function of
  the position of incidence. Two $3\ $mm wide bands
(measured in $y$-direction) of  position of incidence
  have been selected, one crossing the WLS fibers (solid dots),
  the other avoiding the fibers (open squares). In the upper plot the
  FPC surface has been perpendicular to the beam axis while
in the lower plot it has been tilted by $1.7^\circ$.
}
\label{fig:fiber-unif}
\end{figure}

The effect of the enhanced FPC response to particles incident close to a
WLS fiber has been studied using high energy electrons for which
the energy resolution ($\sigma_E / E$) is best. In
Fig.~\ref{fig:fiber-unif} the FPC response to 100 GeV electrons is shown as
a function of the point of incidence. 
The average response is enhanced by
$\sim 15\%$ in regions close to the WLS fibers. This effect is
reduced for runs where the FPC was tilted by $1.7^\circ$
(compare Fig.~\ref{fig:fiber-unif} top and bottom)
corresponding to the mean angle of incidence of beam particles coming
from the ZEUS interaction point.

\subsection{Effect on the energy resolution}
\label{unifwls}

\begin{figure}
\begin{center}
\setlength{\epsfxsize}{13cm}
\hspace*{-0cm}
\epsffile{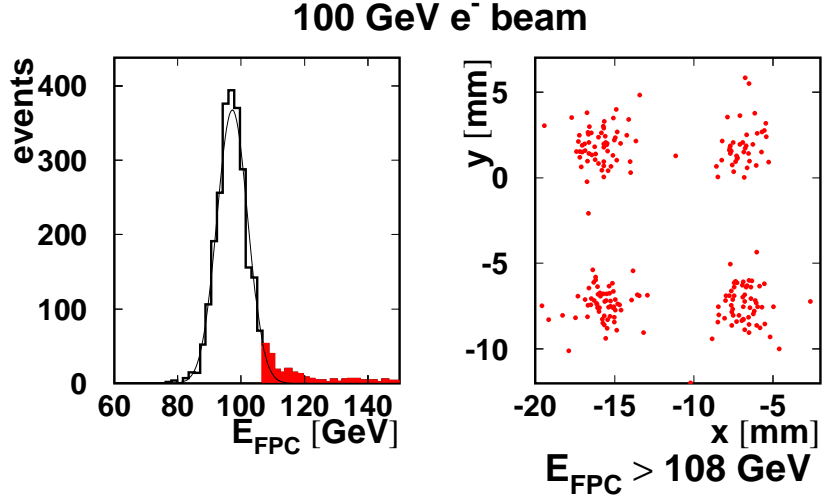}
\end{center}
\vspace*{-1.1 cm}
\caption{\em The left picture shows the distribution of $E_{FPC}$ for 100 GeV
  electrons incident at cell 51. A tail to higher energies can be
  observed (shaded area). In the right picture the 
positions of electrons as measured by the DLWC are shown
for the events in the
  tail.
 These electrons hit the FPC close to one of the four WLS fibers
in the cell.}
\label{fig:tail}
\end{figure}

The enhancement of the signal in the fiber regions also affects the energy
resolution of the FPC as seen in Fig.~\ref{fig:tail}. The
distribution of the total reconstructed energy, $E_{FPC}$, 
shows a tail
to higher energies. Figure~\ref{fig:tail} right shows the positions
of electrons selected from this tail, as measured by the DLWC. These
electrons hit the FPC close to one of the four WLS fibers of the cell.
 The energy
resolution obtained using the 
RMS and the mean value of the distribution ($\bar{E}$)
is RMS/$\bar{E} = 8 \% $, 
while a resolution of 5\% is obtained from a gaussian fit,
which is insensitive to the high energy tail.

\subsection{Tunneling of electrons into the HAC section}
\label{tunneling}

\begin{figure}
\begin{center}
\setlength{\epsfxsize}{13cm}
\hspace*{-0cm}
\epsffile{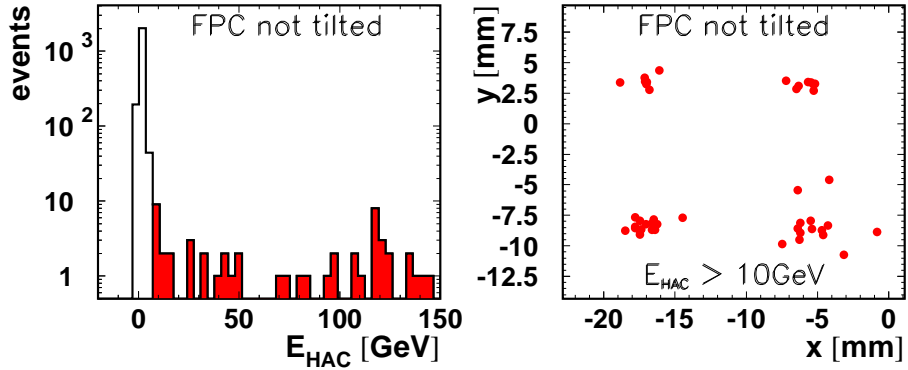}
\end{center}
\vspace*{-1.5 cm}
\caption{\em The energy deposited in the HAC section of
  the FPC by 100 GeV electrons is plotted for vertical incidence.
The plot on the right shows the impact postion of electrons depositing more
  than 10 GeV in the HAC section (shaded area on the left plot).}
\label{fig:el-hac}
\end{figure}

\begin{figure}
\begin{center}
\setlength{\epsfxsize}{8cm}
\hspace*{-0cm}
\epsffile{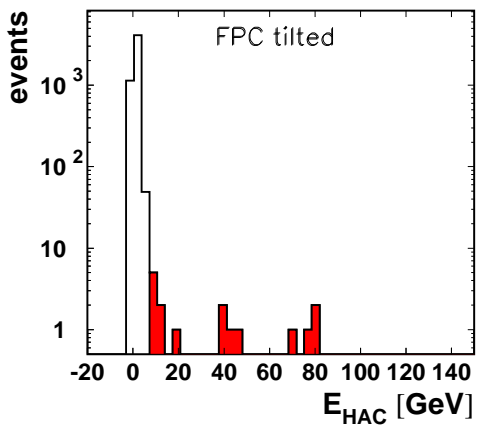}
\end{center}
\vspace*{-1.4 cm}
\caption{\em The energy deposited in the HAC section of
  the FPC by 100 GeV electrons is plotted for the
  FPC tilted by $1.7^\circ$.}
\label{fig:el-hac-tilt}
\end{figure}

For electrons of less than 100 GeV, more than 99\% of their
energy is absorbed in an infinitely wide block of 26.6 $X_0$ depth.
 However, electrons hitting the WLS fibers
 start to shower much later than 
those incident on the lead. Since the fibers
in the EMC section correspond to only $0.4 \, X_0$, electrons
can traverse the EMC section through the fibers and deposit their
energy in the HAC section.
For a data sample where electrons are uniformly distributed in the
transverse plane, about 2\% of the events show more than 10\% of
the incident energy in the HAC section.
These events are also concentrated near the
positions of the WLS fibers
(see Fig.~\ref{fig:el-hac}).
 When the FPC is tilted by $1.7^\circ$ this fraction of electrons
reduces to 0.3\% (see Fig.~\ref{fig:el-hac-tilt}).

\section{FPC performance with electrons}

\subsection{Linearity and energy resolution}
\label{sec-linearityem}

For the study of 
 the linearity and energy resolution, electrons incident 
uniformly on the area of a cell ($24 \times 24\ $mm$^2$)
 have been
selected at all available beam energies.
The total signal has been obtained by summing the signals from the
cluster of $3\times 3$ EMC
cells centered on the cell containing the point of incidence.

\begin{figure}
\begin{center}
\setlength{\epsfxsize}{10cm}
\vspace*{-5.0 cm}
\centerline{\mbox{\epsffile{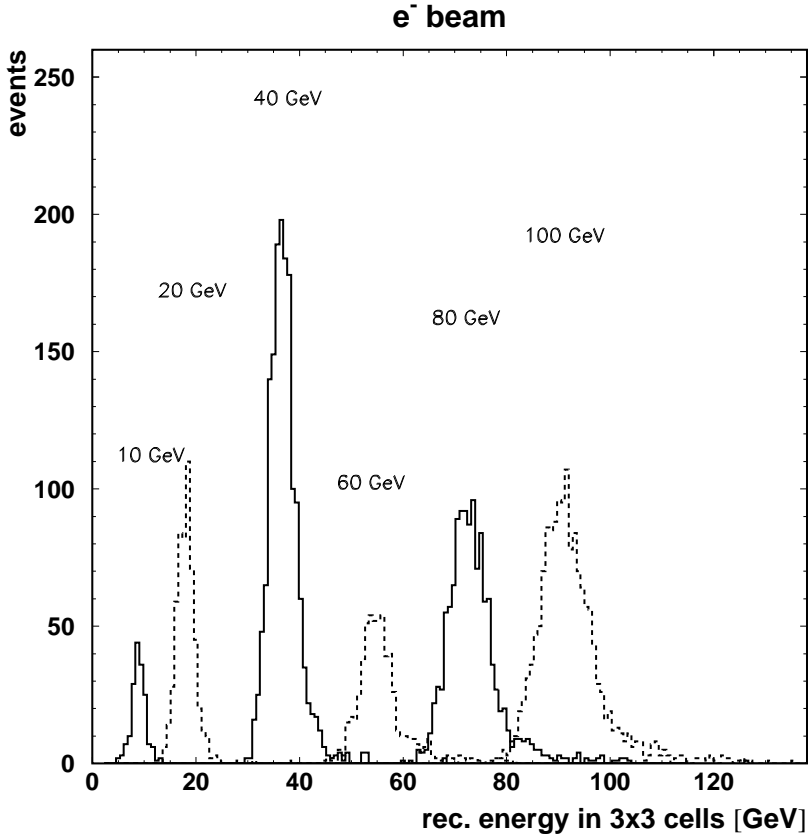}}}
\vspace*{-2.0 cm}
\caption{\em Pulse height signals 
from $3 \times 3$ EMC cells
at different electron beam energies.}
\label{fig:emdistr}
\end{center}
\end{figure}

\begin{figure}
\begin{center}
\setlength{\epsfxsize}{10cm}
\vspace*{-1.0cm}
\centerline{\mbox{\epsffile{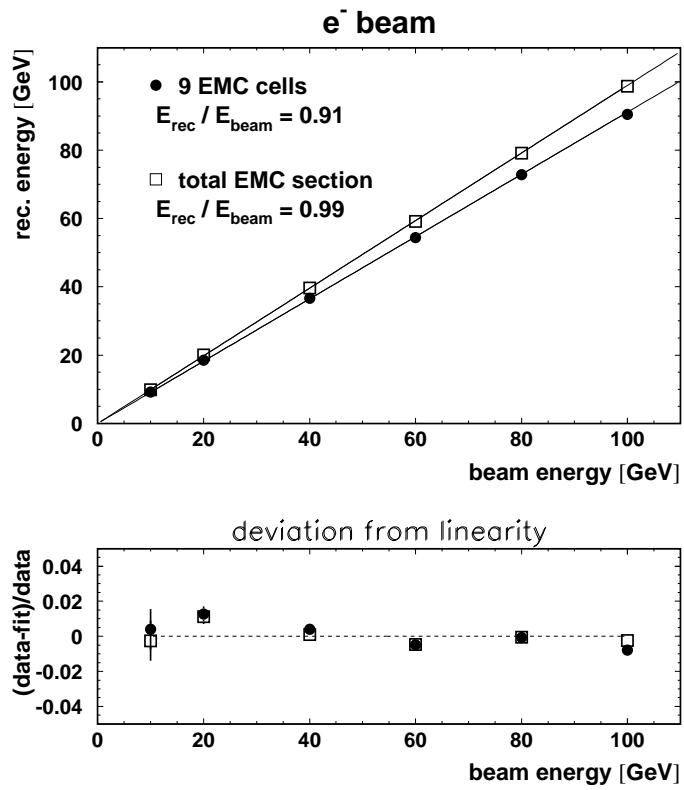}}}
\vspace*{-0.5 cm}
\caption{\em The energy measured for electrons in $3 \times 3$ cells
of the EMC section of FPC and in the complete EMC section. The lower
plot shows the deviations from linearity.}
\label{fig:f2}
\end{center}
\end{figure}

Figure~\ref{fig:emdistr} shows the pulse height distributions for
$3\times 3$ EMC cells at different beam energies. Gaussian fits
have been performed to these distributions.
Figure~\ref{fig:f2} shows the gaussian mean versus beam energy and the
differences from a straight line fit.
Two different cell clusters are considered, namely:
$3\times 3$ EMC cells and the whole EMC section.
About 91\% (99\%) of the beam energy is observed in the cluster
of $3\times 3$ EMC cells (EMC section).
About 1\% of the electron energy leaks to the HAC section.
 The deviations from linearity
are found to be 1\% or smaller for both cluster definitions.

\begin{figure}
\setlength{\epsfxsize}{10cm}
\vspace*{-1.0cm}
\centerline{\mbox{\epsffile{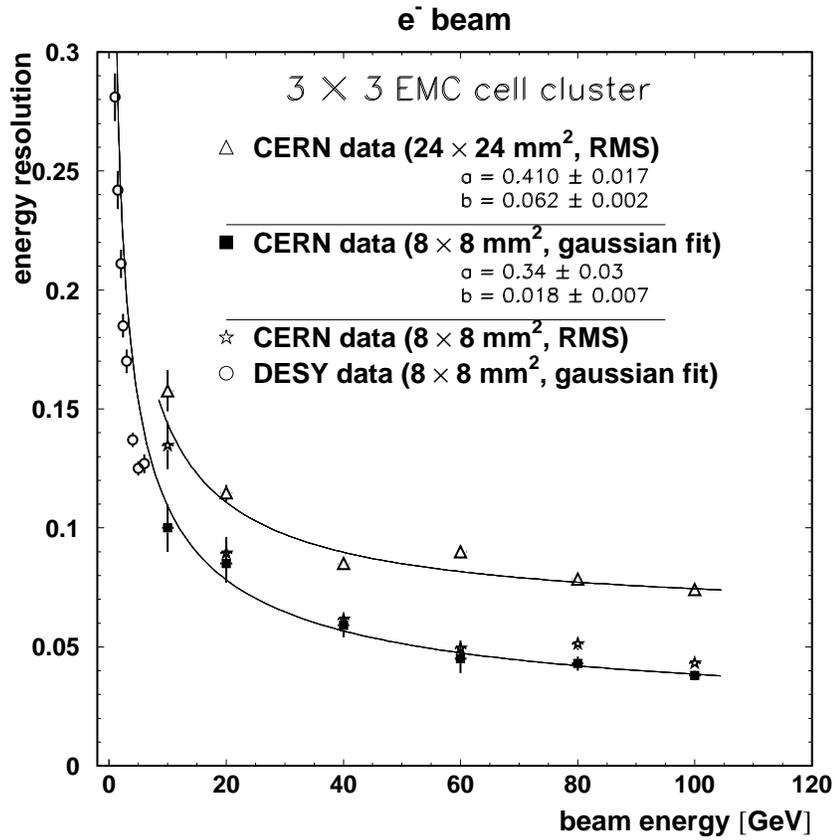}}}
\vspace*{-1.5 cm}
\caption{\em Electron energy resolution obtained from a gaussian fit
to the signals measured with a $3 \times 3$ EMC cell
cluster and from the RMS of the signal distributions.
The beam electrons are distributed either over 
a $24 \times 24\ $mm$^2$ square (cell size)
or over  a $8 \times 8\ $mm$^2$ square.
The curves show  fits to the CERN data.
For comparison also data from the prototype measured at DESY are shown.
}
\label{fig:f10wls}
\end{figure}

 The ratio RMS to mean of the pulse height distribution  
is plotted
versus the beam energy in
Fig.~\ref{fig:f10wls}.
The data are fitted
to the  quadratic sum of a sampling and a constant 
term:

 $$\frac{RMS}{\overline{E}} = \frac{a}{\sqrt{E}} \oplus b$$

where $E$ is in GeV.
The result of the fit is
$ a = (0.410 \pm 0.017)$ GeV$^{1/2}$ and
$ b = 0.062 \pm 0.002 $.

The signal enhancement around the WLS fiber region 
distorts the distribution and
 broadens the resolution as discussed in 
section~\ref{unifwls}.
This can be demonstrated by selecting only
 electrons incident on a square of $8 \times 8\ $mm$^2$
centered at the cell center.
In this way 
the beam impact point is kept away from
 WLS fibers.
Again, the signal obtained from the sum of $3 \times 3$ EMC cells
has been considered.
The relative standard deviations, $\sigma_E / E$,
obtained from gaussian fits
are plotted versus the beam energy in Fig.~\ref{fig:f10wls}. 
The data are fitted
to the  quadratic sum of a sampling and a constant 
term:

 $$\frac{\sigma_E}{E} = \frac{a}{\sqrt{E}} \oplus b$$

where $E$ is in GeV, yielding
$ a = (0.34 \pm 0.03)$ GeV$^{1/2}$,
$ b = 0.018 \pm 0.007 $.
By restriction to the $8 \times 8$ mm$^2$ region on the 
center of the central cell,
a substantial improvement is obtained in the resolution, particularly in
the constant term $b$, which contains the contribution of
non-uniformities.

Lower energy DESY data taken with an FPC prototype 
are included for comparison.
There are mainly two contributions to the measured value $0.34/\sqrt{E}$:
sampling fluctuations and photoelectron fluctuations.
According to MC, the expected sampling contribution to the energy
resolution is $0.29/\sqrt{E}$.
Therefore the contribution from the photoelectron statistics is estimated
as about $0.17/\sqrt{E}$. The LED data taken in those EMC
cells, which have been used for determining
the electron energy resolution,
show photostatistics fluctuations around $0.15/\sqrt{E}$, in good agreement
with the estimate presented above.

\subsection{Position reconstruction}
\label{sec-electronposition}

Position reconstruction 
for incident electrons is done in the following way:

\begin{itemize}
\item
The electromagnetic cell with the maximum energy signal is 
searched for. The event
is accepted if around this cell a $3 \times 3$ EMC
cluster can be defined inside the FPC.

\item
Once the $3 \times 3$ cluster is selected, a linear algorithm is used
to obtain a first approach to the correct position.
In the following only the position recontruction along the $y$-direction
is shown, but the same results are obtained in the $x$-direction.
 The formula applied
is

$$y_{bar} = \frac{\sum_{i=1}^9 S_i \cdot y_i}
{\sum_{i=1}^9 S_i}$$

where $i$ runs over the 9 cells, $y_i$ are the
$y$-coordinates of the
cells and $S_i$ are the energy signals.

\item
The resulting 
$y_{bar}$, when plotted versus the true $y$-coordinate 
as given by the DLWC, 
shows a characteristic
S-shape as shown in Fig.~\ref{fig:f17}.
After applying a suitable
correction to $y_{bar}$ one obtains $y_{rec}$ that, as shown in
Fig.~\ref{fig:f17}, gives an unbiased estimate of the true $y$. The
correction is of the type

$$y_{rec} = P_1 + P_2 \cdot \tanh (P_3 \cdot y_{bar})_{\cdot}$$

\begin{figure}
\setlength{\epsfxsize}{10cm}
\vspace*{-3.0cm}
\centerline{\mbox{\epsffile{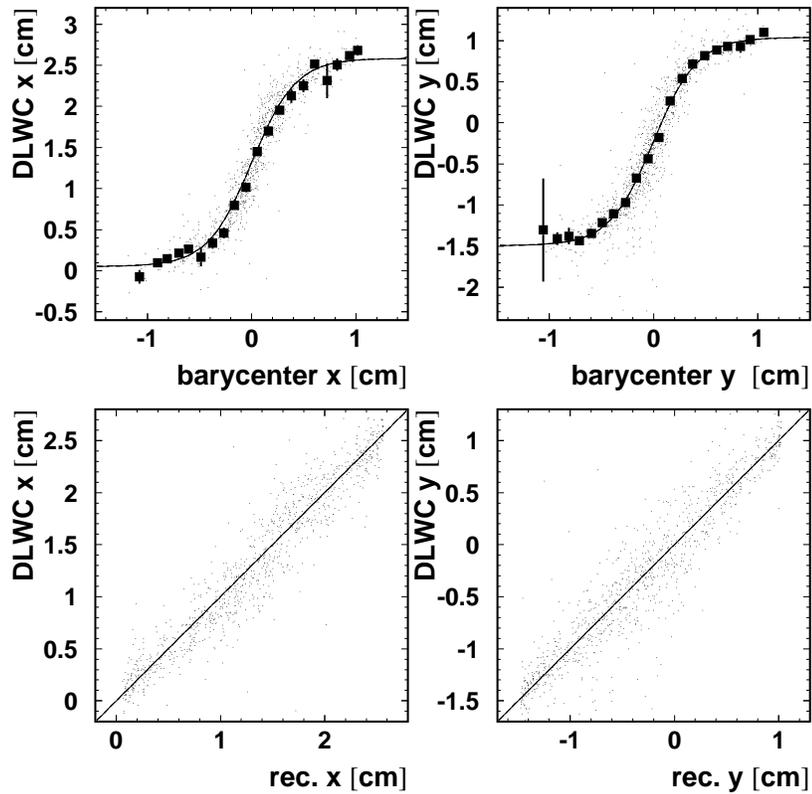}}}
\vspace*{-2.0 cm}
\caption{\em Position reconstruction for 20 GeV electrons
before (upper) and after (lower)
correcting for the S-shape
distortion (see text).}
\label{fig:f17}
\end{figure}

The
correction applied at this stage is approximately independent
of the beam energy, as
one could expect from the fact that the shower transverse profile
is almost 
energy independent \cite{kn:jose}.

\end{itemize}

\begin{figure}
\vspace*{-1.0cm}
\setlength{\epsfxsize}{8cm}
\centerline{\mbox{\epsffile{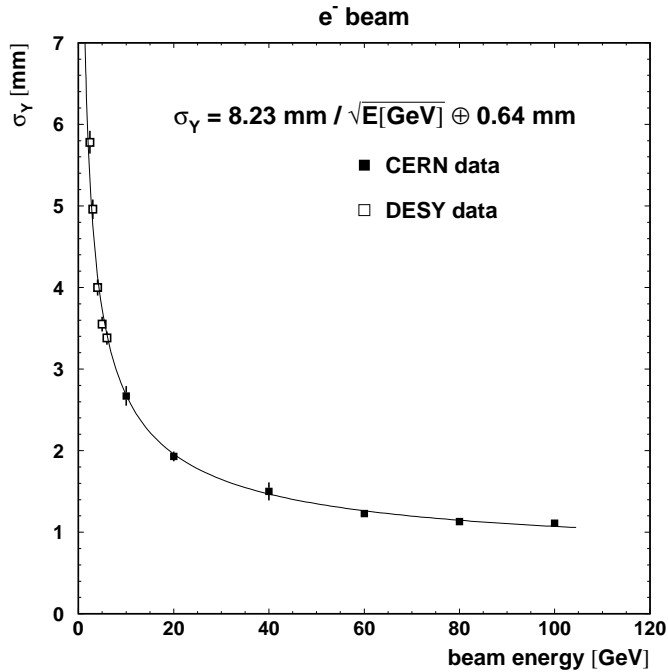}}}
\vspace*{-1.5 cm}
\caption{\em Energy dependence of the position resolution
for electrons.}
\label{fig:f18}
\end{figure}

The 
algorithm for electron position reconstruction
explained above has been applied to 
large data samples taken 
in the CERN and DESY beam tests.
The position resolution is obtained from a gaussian fit to the
$\Delta y$ distribution, where 
$\Delta y = y_{rec} - y_{DLWC}$.
  The
expected $1/\sqrt{E}$ scaling law  is
observed, as can be seen in Fig.~\ref{fig:f18} for the
$y$-coordinate. The same behavior
is seen for $x$.
The result of a fit gives:

$$\sigma_y = 
\frac{0.82\pm 0.01}{\sqrt{E}} \oplus 0.064\pm0.005 \,\,
\mathrm{cm}$$

where $E$ is measured in units of GeV.

\section{FPC performance with pions}
\label{sec-pions}

\subsection{Energy response}

The $\pi^-$ beam provided by the CERN test beam facility showed
a substantial muon contamination. These muons have been rejected from
the data sample by applying the following cuts:

\begin{itemize}
\item $E_{b5} < 0.25\ $mip
\item $E_{FPC} > 0.25 \cdot E_{beam}$
\end{itemize}

where 
$E_{b5}$ is the signal measured in counter $b_5$
(see Fig.~\ref{fig:cerntest_engl}),
1 mip is the most probable 
signal deposited by a minimum ionizing particle
and $E_{FPC}$ is the total energy measured in the FPC.

\begin{figure}
\begin{center}
\setlength{\epsfxsize}{10cm}
\vspace*{-3.0 cm}
\centerline{\mbox{\epsffile{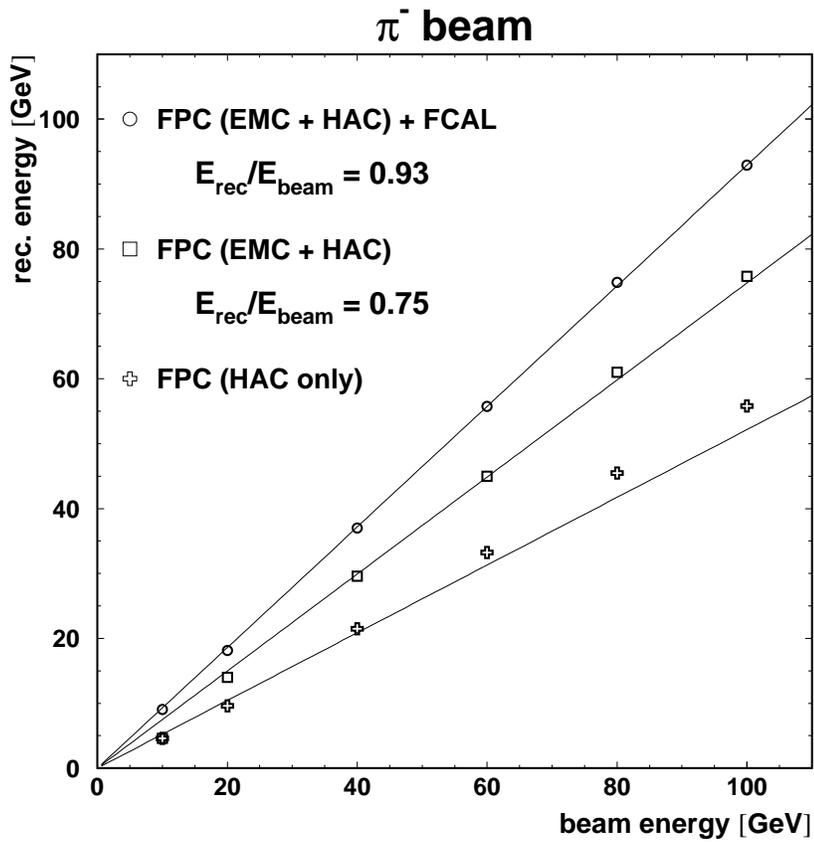}}}
\vspace*{-5.0 cm}
\setlength{\epsfxsize}{10cm}
\centerline{\mbox{\epsffile{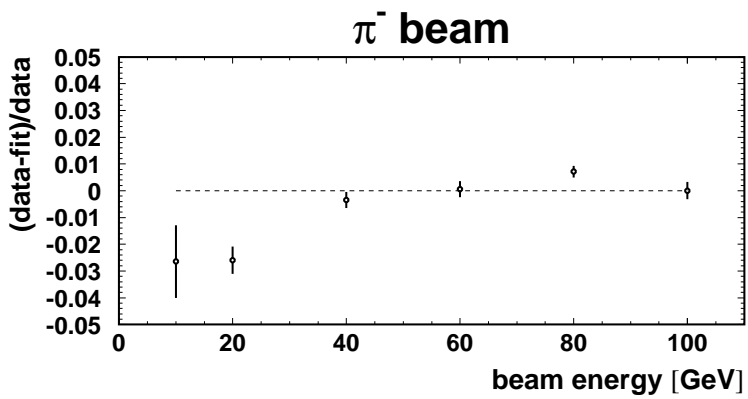}}}
\vspace*{-4.8 cm}
\caption{\em 
The energy measured for $\pi^-$ in the HAC section of FPC,
in the FPC  and 
in the FPC $+$ FCAL prototype modules.
The lower plot shows the 
deviations from
linearity for the total FPC $+$ FCAL prototype signal.}
\label{fig:f32}
\end{center}
\end{figure}

The $\pi^-$ signal has been measured with different
combinations of the FPC and FCAL prototype modules. The FCAL
prototype modules are $7 \, \lambda$ deep. For this
beam test the modules have been calibrated with electrons and
muons. Previously, a setup of four FCAL prototype modules
had been tested with electrons, muons and hadrons, and the
calorimeter was found to be compensating, $e/h = 1.0$ for 
momenta $p \geq 3$ GeV to within 3\%~\cite{kn:FCAL}.

Figure~\ref{fig:f32} top shows the signals measured with the
FPC and FCAL prototype as a function of the $\pi^-$ beam
energy.
       For the FPC alone and FPC+FCAL prototype the measured signal
        follows closely a linear rise with the beam energy. For
        FPC+FCAL prototype, the deviations
from  linearity are $<$3\% (see Fig.~\ref{fig:f32} bottom).

Energy fractions of about 56\% (FPC HAC), 75\% (FPC: EMC+HAC)
and 93\% (FPC + FCAL prototype) are measured at 
$E_{\pi} = 100$ GeV.
These fractions change slowly with beam energy.

\begin{figure}
\setlength{\epsfxsize}{10cm}
\vspace*{-3.0 cm}
\centerline{\mbox{\epsffile{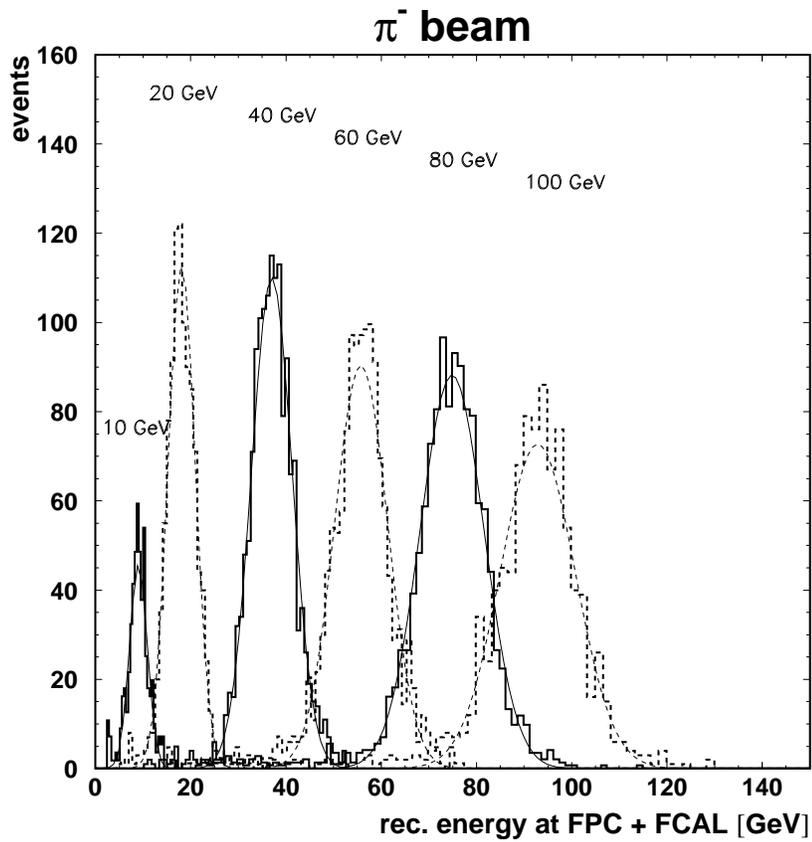}}}
\vspace*{-2.0 cm}
\caption{\em Pulse height signals observed in the FPC $+$ FCAL for 
$\pi^-$ beam
energies of 10, 20, 40, 60, 80 and 100 GeV.}
\label{fig:f31_1}
\end{figure}

Signal distributions
obtained when combining the signals from FPC and FCAL prototype
are shown in Fig.~\ref{fig:f31_1}
at beam energies between 10 and 100 GeV.

\subsection{Energy resolution}

\begin{figure}
\setlength{\epsfxsize}{8cm}
\vspace*{-0.0 cm}
\centerline{\mbox{\epsffile{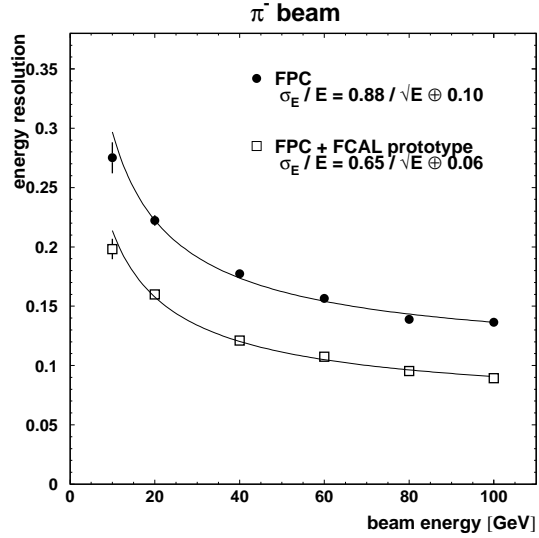}}}
\vspace*{-0.7 cm}
\caption{\em 
Energy resolution ($\sigma_E / E$) for $\pi^-$ as measured by 
FPC alone and
FPC + FCAL prototype as a function of the 
$\pi^-$ beam energy.
}
\label{fig:f33}
\end{figure}

Figure~\ref{fig:f33} shows the 
energy resolution for $\pi^-$ as a function of energy
for FPC alone and FPC + FCAL prototype.
Each point is the result of a gaussian fit to the 
corresponding
$\pi^-$ distribution shown in Fig.~\ref{fig:f31_1}.
 The curve shows a
fit  with the
parametrization

$$\frac{\sigma_E}{E} = \frac{a}{\sqrt{E}} \oplus b$$

where $E$ is in GeV,
yielding:
$a = (0.65 \pm 0.02)$ GeV$^{1/2}$ ; $b = 0.06 \pm 0.01$.

The
resolution is affected by transverse leakage
(see  Fig.~\ref{fig:cern_engl}).
When installed in
  ZEUS the FPC is completely surrounded
by FCAL modules.
 As a result there is no transverse leakage except into the
beam hole and,
according to MC,
the energy resolution in ZEUS improves by  20\%.

\subsection{Position resolution}
\label{sec-pionposition}

The position reconstruction for $\pi^-$ has been studied using
three different sets of cells corresponding to the
FPC-EMC section, the FPC-HAC section and the combined
EMC $+$ HAC sections.
In the following only the position recontruction along the $y$-direction
is shown, but the same results are obtained in the $x$-direction.
A barycenter of the cell coordinates is determined as

$$y_{bar} = \frac{\sum_{i} w_i \cdot y_i}
{\sum_{i} w_i}$$

where $i$ runs over all cells considered,
$y_i$ is the $y$-coordinate of the center of cell $i$,
        and $w_i$ are weights
defined as 

$$w_i = \max \left( 0, \ln \left( S_i/\sum_i S_i \right) - c_i \right)$$

$S_i$ is the energy signal in cell $i$ and $c_i$ is the
threshold in the quantity 
$\ln \left( S_i/\sum_i S_i \right)$
corresponding to cell $i$ below which the cell is not included in the
barycenter calculation.

The relation between the position $y$ given by the DLWC ($y_{DLWC}$)
and the 
$y_{bar}$ measured in the FPC shows a negligible S-shape distortion.
This is due to the use of a logarithmic function of the normalized
signal to determine the weights $w_i$.
Hence no S-shape correction is needed and 
the reconstructed position
$y_{rec}$
 has been obtained by just rescaling the
barycenter $y_{bar}$.

\begin{figure}
\setlength{\epsfxsize}{12cm}
\epsffile{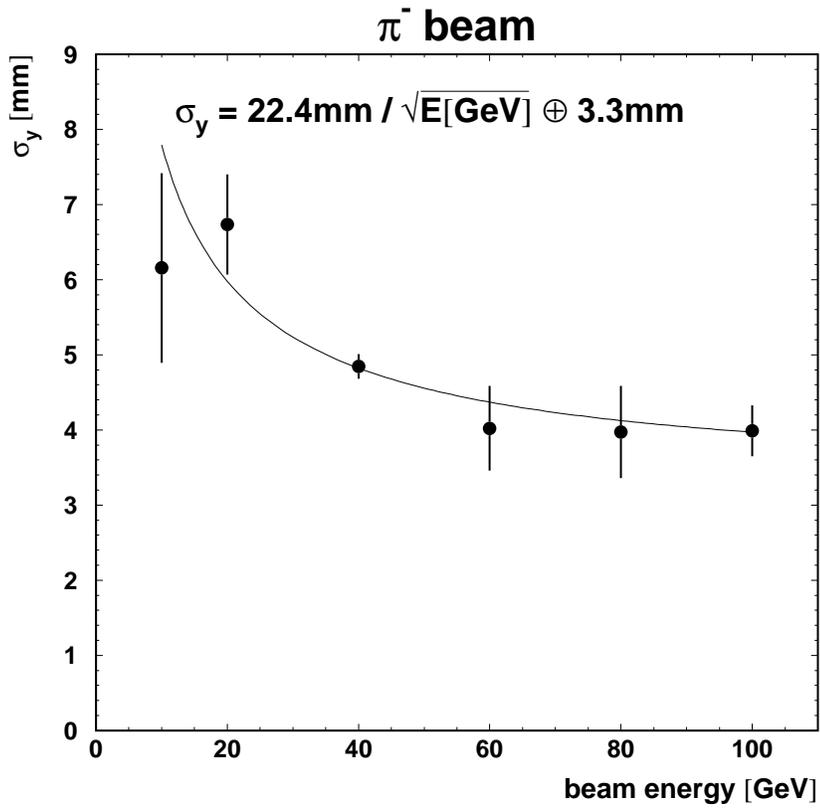}
\vspace*{-1.0 cm}
\caption{\em The $\pi^-$
position resolution, using EMC and HAC section
cells, as a function of the beam energy. The resolution is
calculated as the $\sigma$ of a gaussian fit to the difference
between reconstructed and true (DLWC) position.}
\label{fig:pion_pos_resol}
\end{figure}

The position resolution is obtained from a gaussian fit to the
$\Delta y$ distribution, where 
$\Delta y = y_{rec} - y_{DLWC}$.
Figure~\ref{fig:pion_pos_resol} shows the measured resolution,
$\sigma_y$, as a function of the beam energy.
All the cells of the FPC have been used to determine the 
barycenter.
The data are fitted using:

$$\sigma_y = \frac{a}{\sqrt{E}} \oplus b$$

where $E$ is in GeV.
The result of the fit yields:
 $a = 22 \pm 3 $ mm$\cdot$GeV$^{1/2}$
 and $b = 3.3 \pm 0.5 $ mm.
When only EMC cells or only HAC cells are used to obtain the
barycenter, the resolution degrades as shown in
Table~\ref{tablita}.
For the set of EMC cells, only events which deposit more than 20\%
of the energy in the EMC section are considered
and the RMS is used instead
of the sigma from the gaussian fit.

\begin{table}
\begin{tabular}{|c|c|c|}
\hline
{\bf Cells used} & $a \, ($mm$ \cdot $GeV$^{1/2})$ & $b \, ($mm$)$ \\
\hline \hline
EMC $+$ HAC ($\sigma$) & $22 \pm 3$   &  $3.3 \pm 0.5$ \\
\hline
HAC ($\sigma$)         & $50 \pm 4$   &  $2 \pm 3$ \\
\hline
EMC ($E_{EMC} > 0.2 \cdot E_{beam}$, RMS) & $74 \pm 8$ & $9.6 \pm 0.9$ \\
\hline
\end{tabular}
\caption{\em $\pi^-$ position resolution using different sets of cells,
with the $\sigma$ from a gaussian fit or the RMS,
parametrized as the quadratic sum of a sampling and
a constant term.}
\label{tablita}
\end{table}

\section{Monte Carlo for the ZEUS environment}

The FPC has been implemented in the Monte Carlo program MOZART
for the simulation of the ZEUS detector response.
The program MOZART uses the general purpose packages
GEANT 3.13 and GHEISHA  modified to describe
the test beam data measured with the ZEUS uranium calorimeter
modules.
In particular, for the $e/h$ ratio and energy resolution of hadrons,
discrepancies in the range 20-30\% were found
between the GEANT 3.13 + GHEISHA prediction and 
the measurements from the FCAL prototype \cite{kn:hartner}.
To overcome these discrepancies a shower terminator was introduced
in the GEANT 3.13 + GHEISHA package, which operates on the
evaporation energy of excited nuclei and on neutrons with 
kinetic energies below 50 MeV \cite{kn:sim-zeus}.
This shower terminator had free parameters which were chosen  such
that the discrepancies between Monte Carlo prediction and test beam
data were below 5\%.
In addition, and because the package  GEANT 3.13 was very time 
consuming by tracking shower particles in the calorimeter,
a new shower terminator was introduced which
operates on $e^-$, $e^+$ and $\gamma$ with kinetic energies below
200 MeV \cite{kn:sim-zeus}.
Using both shower terminators in the GEANT 3.13 + GHEISHA package
the execution time was reduced by a factor of about 15.

\begin{figure}
\setlength{\epsfxsize}{10cm}
\vspace*{-0.0 cm}
\epsffile{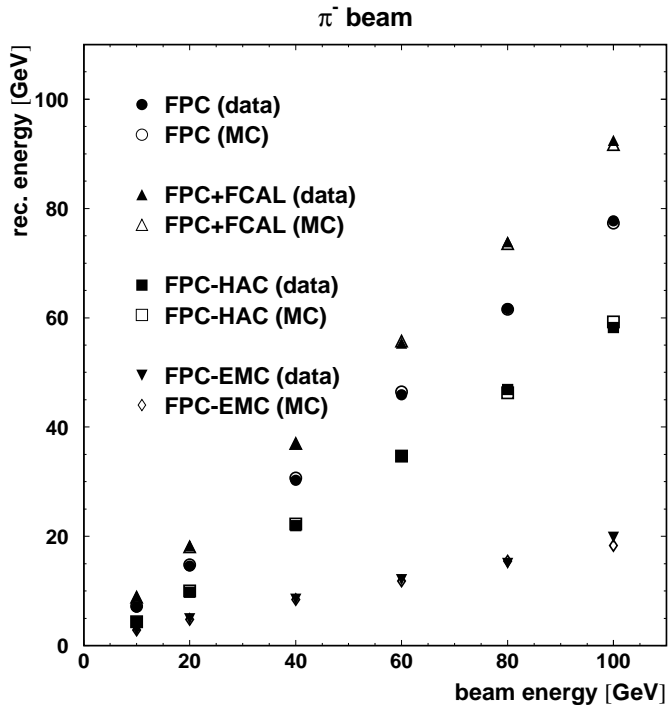}
\vspace*{-0.0 cm}
\caption{\em
The mean of the reconstructed energy in FPC, FPC+FCAL, FPC-HAC
  and FPC-EMC for $\pi^-$ is shown as a function of the beam energy.
The results of the Monte Carlo for the CERN-test geometry are shown in
comparison with the data.
}
\label{fig:mozart_lin_pi}
\end{figure}

\begin{figure}
\setlength{\epsfxsize}{10cm}
\vspace*{-0.0 cm}
\epsffile{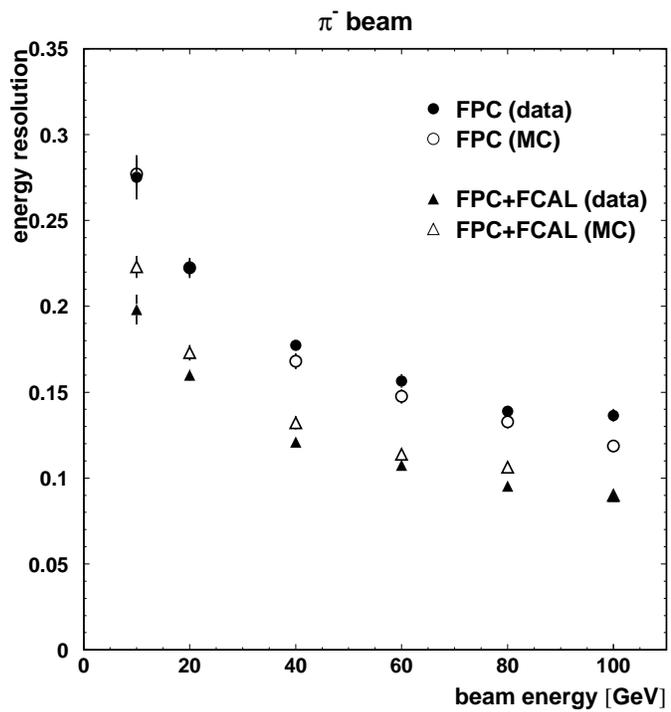}
\vspace*{-0.0 cm}
\caption{\em
The energy resolution for $\pi^-$ in FPC and FPC+FCAL
prototype is shown
  as a function of the beam energy.
The results of the Monte Carlo for the CERN-test geometry are shown in
comparison with the data.
}
\label{fig:mozart_reso_pi}
\end{figure}

The scheme of shower terminators has also been used to simulate
the FPC response in 
GEANT 3.13 + GHEISHA.
Since the FPC and FCAL are different calorimeters,
a different set of parameters of the shower terminators has been 
chosen for the shower particles inside the FPC.
In Fig.~\ref{fig:mozart_lin_pi}  average energies deposited in
the different sections 
(FPC-EMC, FPC-HAC, FPC and FPC + FCAL prototype)
are shown as a function of the beam energy for 
test beam data and MOZART.
Good agreement for all sections is observed.
Energy resolutions (from gaussian fits) are plotted in
Fig.~\ref{fig:mozart_reso_pi}
as a function of the beam energy for data and MOZART.
The FPC energy resolution is reasonably well described.
When adding the FCAL prototype signal to the FPC signal
small discrepancies of the order of 5-10\% are found.

\section{Conclusions}
\label{sec-conclusions}

The Forward Plug Calorimeter (FPC)
for the ZEUS detector
at the electron-proton collider HERA at DESY, 
has been tested at DESY and CERN with beams
of electrons, muons and pions in the
range 1 to 100 GeV in 1997. The results of these tests have been used
to determine the calibration of the FPC.
The energy and position resolutions, $\sigma_E$ and $\sigma_y$
respectively, 
measured for both electrons and
$\pi^-$ can be summarized as follows:

\begin{tabular}{ll}
for electrons: &
$\sigma_E / E = (0.41\pm 0.02)/\sqrt{E} \oplus 0.062 \pm 0.002$
\\
&
$\sigma_y  = (0.82\pm 0.01)/\sqrt{E} \oplus 0.064\pm0.005 \,\,$cm
\\
&  \\
for pions: &
$\sigma_E / E = (0.65\pm 0.02) / \sqrt{E} \oplus 0.06\pm 0.01\%$
\\
&
$\sigma_y  = (2.2\pm 0.3) / \sqrt{E} \oplus 0.33\pm 0.05 \,\,$cm
\\
\end{tabular}

where $E$ is measured in units of GeV.

A monitoring
system using a $^{60}$Co source has been
used to transport the calibration constants from the test 
beam to the
ZEUS detector and to monitor the stability of the calibration
since its installation in 1998.

\vspace*{1 cm}

\flushleft{\Large \bf Acknowledgements}

We would like to thank the staff support from the various 
Institutes which collaborated in the construction of the FPC and
in the setup of the test systems, in particular
J. Hauschildt and  K. L\"offler (DESY),
R. Feller, E. M\"oller and H. Prause (I. Inst. for Exp. Phys., Hamburg),
A. Maniatis (II. Inst. for Exp. Phys., Hamburg)
and 
the members of the mechanical workshop of the Faculty of Physics 
from Freiburg University.
We also would like to thank L. Herv\'as (CERN) for helping us with
the readout electronics during the beam test.
We are grateful for the Weizmann group, in particular to
Prof. Y. Eisenberg, for the support in the early stages of the 
project.
We are grateful for the hospitality of CERN and for the support of
the CERN technical staff during the measurements.

\end{document}